\documentclass[aps,prd,showpacs,twocolumn,floatfix]{revtex4}
\usepackage{graphicx}% Include figure files
\usepackage{dcolumn}% Align table columns on decimal point
\usepackage{bm}% bold math
\usepackage{feynmf}%%%{feynmp}
\begin{document}

\def\be{\begin{equation}}
\def\ee{\end{equation}}
\def\bd{\begin{displaymath}}
\def\ed{\end{displaymath}}
\def\ba{\begin{eqnarray}}
\def\ea{\end{eqnarray}}
\def\J{\rm J}
\def\P{\rm P}
\def\C{\rm C}
\def\x{x}
\def\y{y}

\title{\Large 
Associated Charmonium Production in Low Energy $p\bar{p}$ Annihilation\\
} 
\author{
T.Barnes$^{a,b}$\footnote{Email: tbarnes@utk.edu}
and
X.Li$^{b}$\footnote{Email: xli22@utk.edu}
}
\affiliation{
$^a$Physics Division, Oak Ridge National Laboratory,
Oak Ridge, TN 37831, USA\\
$^b$Department of Physics and Astronomy, University of Tennessee,
Knoxville, TN 37996, USA
}

%\date{\today}

\begin{abstract}
The QCD mechanisms underlying the exclusive strong decays and hadronic 
production amplitudes of charmonium remain poorly understood, despite 
decades of study and an increasingly detaled body of experimental information. 
One set of hadronic channels of special interest 
are those that include baryon-antibaryon states. These are 
being investigated experimentally at BES and CLEO-c 
in terms of their baryon resonance 
content, and are also of interest for the future PANDA 
experiment, in which charmonium and charmonium hybrids will be produced
in $p\bar p$ annihilation  
in association with light mesons. 
In this paper we develop a simple initial-state light meson emission model 
of the near-threshold 
associated charmonium production processes 
$p\bar p \to \pi^0 \Psi$, 
and evaluate the differential and total cross sections 
for these reactions
in this model. 
(Here we consider the states $\Psi = \eta_c, J/\psi, \psi', \chi_0$ and
$\chi_1$.) 
The predicted near-threshold cross section for $p\bar p \to \pi^0 J/\psi$ 
is found to be numerically similar to two previous theoretical  
estimates, and is roughly comparable to the (sparse) existing data for this process. 
The theoretical
charmonium angular distributions predicted by this model are far from 
isotropic, 
which may be of interest for PANDA detector design studies. 
\end{abstract}


\maketitle

\section{Introduction}
Strong decays of charmonia through annihilation of the $c\bar c$ pair to
light hadrons dominate the total widths of the lighter charmonium states \cite{PDG}.
In contrast to the faster open-charm decays, which appear to be reasonably 
well described by both the $^3$P$_0$ model \cite{Barnes:2005pb} and the 
Cornell timelike vector decay model 
\cite{Eichten:2004uh} (given the current state of the data),
the $c\bar c$ annihilation decays are much less well understood. There are 
several general 
rules that appear to be respected by these decays; in particular, the number of
QCD vertices in the leading-order Feynman diagram for annihilation into
light quarks and gluons is a useful guide. For example, the C = $(+)$ charmonia 
that can annihilate at $O(g^2)$ (into $gg$ for 
$\eta_c, \chi_0, \chi_2$, and 
$q\bar q g$ for the $\chi_1$) have strong annihilation widths 
of $\sim 1-25$~MeV, often
much larger than the $\sim 0.1-0.3$~MeV widths of the 
C = $(-)$ states $J/\psi$ and 
$\psi'$, which must annihilate at $O(g^3)$ into $ggg$. Since $c\bar c$ 
annihilation is a short-ranged process (the charm quark propagator implies 
a range of $r\sim 1/m_c$), a strong suppression of annihilation widths with 
increasing orbital angular momentum L$_{c\bar c}$ is also anticipated; 
this suggests for example that the D-wave $c\bar c$ states
$^1$D$_2$ and $^3$D$_2$, if below their DD* open-charm threshold,
will have strong widths of $< 1$~MeV \cite{Eichten:2004uh,Barnes:2003vb}.

Although the inclusive annihilation decays are qualitatively understood 
in terms of $c\bar c$ annihilation into gluons,
exclusive $c\bar c$ annihilation
decays remain a mystery, despite the existence of considerable experimental 
information on the branching fractions of some $c\bar c$ states into specific
exclusive modes. In particular, much is known about the exclusive two-body 
annihilation decays of the $J/\psi$ and $\psi'$, and a ``12\% rule" for the 
relative branching fractions of the $\psi'$ and $J/\psi$ into many of these modes 
is part of charmonium folklore. The recent increase in the number of modes 
studied, for example by CLEO-c~\cite{Briere:2005rc}, has made it clear however that this rule is 
not generally applicable. 

In this paper we note that there may be a simple relation between some two-body 
and three-body annihilation decays of charmonia, specifically in decays to 
the final states $p\bar p$ and $p\bar p \pi^0$. These decays are of interest 
both as a novel technique for studying N$^*$ spectroscopy, for example 
at BES \cite{Bai:2001ua,Ablikim:2004ug,Ablikim:2005ir,Liang:2004sd}
and because they can be used to estimate the cross sections 
for associated charmonium production in $p\bar p$ annihilation, as in 
$p\bar p \to \pi^0 + \Psi$ \cite{Lundborg:2005am}.
(We use $\Psi$ to represent a generic charmonium state.) 
These cross sections are of particular interest 
in that they will be exploited by the PANDA project at GSI 
\cite{PandaTechnicalProgress} to search for excited charmonia and charmonium 
hybrids. The scales of these cross sections are at present 
largely unknown; a better understanding of these 
associated charmonium production processes near threshold 
is obviously crucial for planning 
various aspects of this experiment, such as detector design and 
data acquisition.

\section{The Model}

\subsection{Motivation}

Since strong decays to three-body final states are typically dominated by 
quasi-two-body transitions, one might anticipate that decays of the type 
$\Psi \to p\bar p \pi^0$ could be described as two-step processes, 
$\Psi \to ({\rm N}^{*+}\bar p + h.c.) \to  p\bar p\pi^0$, where the 
important N$^*$ baryon isobars are those with large N$\pi$ couplings. 
Specifically we might expect the nucleon itself to make a large or perhaps 
dominant contribution, in view of the large NN$\pi$ coupling. This in turn 
suggests that the associated production of a $\Psi$ state and a pion in 
$p\bar p$ annihilation may take place through initial emission of a pion 
from the incoming $p$ or $\bar p$ line, followed by direct annihilation of 
the $p\bar p$ state into $c\bar c$. At tree level this process is described 
by the two Feynman diagrams of Fig.\ref{fig:diags}.

\begin{center}
\begin{figure}[h]
\vskip 0.5cm
\begin{fmffile}{fmf_diags}
\fmfset{arrow_len}{3mm}
\begin{fmfgraph*}(25,12)
\fmfleft{pbar,p} 
\fmflabel{\large $p$}{p} 
\fmflabel{\large $\bar p$}{pbar}
\fmfright{psi,dum,pi0} 
\fmflabel{\large $\pi^0$}{pi0} 
\fmflabel{\large $\Psi$}{psi}
\fmf{fermion}{p,V1,V2,pbar}
\fmf{phantom}{V1,pi0}
\fmf{phantom}{V2,psi}
\fmfdot{V1,V2}
\fmflabel{\hskip 2.5mm {\bf +}}{dum} 
\fmffreeze
\fmf{dashes}{V1,pi0}
\fmf{boson} {V2,psi}
\end{fmfgraph*}
\begin{fmfgraph*}(10,12)
\end{fmfgraph*}
\begin{fmfgraph*}(25,12)
\fmfleft{pbar,p} 
\fmflabel{\large $p$}{p} 
\fmflabel{\large $\bar p$}{pbar}
\fmfright{psi,dum,pi0} 
\fmflabel{\large $\pi^0$}{pi0} 
\fmflabel{\large $\Psi$}{psi}
\fmf{fermion}{p,V1,V2,pbar}
\fmf{phantom}{V1,pi0}
\fmf{phantom}{V2,psi}
\fmfdot{V1,V2}
\fmffreeze
\fmf{dashes}{V2,pi0}
\fmf{boson} {V1,psi}
\end{fmfgraph*}
\end{fmffile}
\vskip 0.5cm
\caption{Feynman diagrams assumed in this model of 
the generic reaction $p\bar p \to \pi^0 \Psi$.}
\label{fig:diags}
\end{figure}
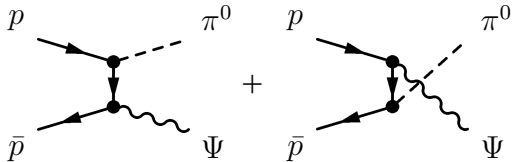
\end{center}
\vskip -1cm

The use of these simple hadron-level ``pole" diagrams to describe 
these processes was previously suggested by Gaillard, Maiani and Petronzio
\cite{Gaillard:1982zm}, who considered the production of charmonium from 
$p\bar p$ states with specified initial quantum numbers. 
Their work was subsequently extended 
analytically by Lundborg, Barnes and Wiedner 
\cite{Lundborg:2005am} (much of Ref.\cite{Gaillard:1982zm} was numerical).
Lundborg {\it et al.} also showed that a constant amplitude approximation 
could be used to estimate the cross section for $p\bar p \to \pi^0 \Psi$ from
the corresponding three-body decay $\Psi \to p\bar p \pi^0$; numerical results
for $\sigma(p\bar p \to \pi^0 J/\psi)$ using both approaches are given in that 
reference. 

The work presented here gives extensive analytic results 
for differential and total cross sections derived from the pole model, 
assuming plane wave initial $p\bar p$ states.  
If the partial width of a charmonium state $\Psi$ to $p\bar p$ is 
known experimentally, one can use this to estimate the corresponding 
$\Psi - p\bar p$ coupling constant $g_{p\bar p\Psi}$. Since the 
NN$\pi$ coupling constant is known, one may then evaluate the differential 
and total cross sections for $p\bar p \to \pi^0 \Psi$ both analytically
and numerically in this model. Here we carry out this exercise for 
the cases of five low-lying $\Psi$ states with known $p\bar p$ partial 
widths, specifically the $\eta_c, J/\psi, \psi', \chi_0$ and $\chi_1$.

\subsection{Amplitudes}

We describe the reactions $p\bar p \to \pi^0 \Psi$ 
as two step processes,
initial pion emission from an incident proton or antiproton followed by 
direct annihilation of the proton-antiproton pair, $p\bar p \to \Psi$. 
This process
at tree level is described by the two Feynman diagrams of Fig.1.
The $pp\pi^0$ and $\bar p \bar p\pi^0$ vertices are taken to be the usual
$g_{pp\pi} \gamma_5$, and the $p\bar p\Psi$ vertex is generically
$g_{p\bar p \Psi} \Gamma$, where the Dirac matrix $\Gamma$ 
specifies the quantum numbers
of the state $\Psi$. We use 
$\Gamma = \gamma_5$ for the $\eta_c$,  
$-i$ for the $\chi_0$,
$-i\gamma_{\mu}$ for the $J/\psi$ and $\psi'$, and
$-i\gamma_{\mu}\gamma_5$ for the $\chi_1$. 
For the vector and axial-vector cases $J/\psi, \psi'$ and
$\chi_1$, $\Gamma$ 
is implicitly contracted into the final 
$\Psi$ polarization four-vector $\epsilon^*_{\mu}$. 

The resulting invariant amplitude is
\be
{\cal M} 
= ig_{\pi} g_{\Psi} 
{\bar v}_{\bar p \bar s}
\Big[\, 
\Gamma\,
\frac{(\slash{\hskip-2mm p} - \slash{\hskip-2mm k} + m)}{(t-m^2)}\,  
\gamma_5
+
\gamma_5\,
\frac{(\slash{\hskip-2mm k} - \slash{\hskip-2mm {\bar p}} + m)}{(u-m^2)}\,  
\Gamma\,  
\Big]
u_{ps}\; .
\label{eq:Minvar_1}
\ee
Here and in the following expressions $m$ is the proton mass, 
$m_{\pi}$ is the pion mass,
$M$ is the mass of the charmonium state $\Psi$, and 
$r_{\pi} = m_{\pi}/m$ and $r_{\Psi} = M/m$ 
are dimensionless mass ratios relative to the proton. 
The pion-nucleon coupling constant is 
$g_{\pi} \equiv g_{pp\pi}$,
and the (state-dependent) charmonium-$p\bar p$ coupling constant is 
$g_{\Psi} \equiv g_{p\bar p\Psi}$. 
We also define 
squared strong coupling constants
$\alpha_{\pi} \equiv g_{pp\pi}^2/4\pi$ and
$\alpha_{\Psi} \equiv g_{p\bar p\Psi}^2/4\pi$. 

It is sometimes useful to rewrite this invariant amplitude in an equivalent 
form that makes the overall $t\leftrightarrow u$ crossing symmetry more evident; 

\bd
\hskip -3cm
{\cal M} 
= 
-\frac{ig_{\pi}g_{\Psi}/2}{(t-m^2)(u-m^2)}
\; \cdot
\ed
\be
\bigg[
(s-M^2 - m_{\pi}^2)\;
{\bar v}_{\bar p \bar s}
\{ 
\gamma_5\slash{\hskip-2mm k},
\Gamma
\}
u_{ps}
- 
(t-u)\
{\bar v}_{\bar p \bar s}
[\,
\gamma_5\slash{\hskip-2mm k}, 
\Gamma\, 
]
u_{ps}\
\bigg]\ .
\ee
Since the commutator (anticommutator) of 
$\gamma_5\slash{\hskip-2mm k}$ with $\Gamma$
vanishes for scalar (pseudoscalar) $\Psi$, 
this rearrangement simplifies
the calculation considerably in these cases.

\subsection{Massless pions}

\subsubsection{Differential cross sections}

The differential and total cross sections we find given this invariant
amplitude are rather simple in the massless pion limit, and only involve two 
independent functions. For this reason we first give results for massless 
pions, and then treat each set of charmonium quantum numbers separately 
for nonzero pion mass.
Our results for the unpolarized differential cross sections
in the massless pion limit are 

\be
\hskip -0.5cm
\langle
\frac{d\sigma}{dt}
\rangle
\Big|_{p\bar p \to \pi^0 \eta_c} 
= 
\frac{\pi}{2}
\frac{\alpha_{\pi}\alpha_{\Psi}}{s (s-4m^2)}\,
\frac{(\x-\y)^2}{\x\y}\ , 
\label{eq:dsigdt1} 
\ee

\bd
\hskip -0.7cm
\langle
\frac{d\sigma}{dt}
\rangle
\Big|_{p\bar p \to \pi^0 \chi_0} 
=
\frac{1}{2}
\langle
\frac{d\sigma}{dt}
\rangle
\Big|_{p\bar p \to \pi^0 (J/\psi, \psi')} 
=
\ed
\be 
\hskip +1.1cm
\frac{\pi}{2}
\frac{\alpha_{\pi}\alpha_{\Psi}}{s (s-4m^2)}\,
\frac{f^2}{\x\y} \ ,
\label{eq:dsigdt2} 
\ee

\bd
\hskip -2.1cm
\langle
\frac{d\sigma}{dt}
\rangle
\Big|_{p\bar p \to \pi^0 \chi_1} 
= 
\pi
\frac{\alpha_{\pi}\alpha_{\Psi}}{s (s-4m^2)}\, \cdot
\ed
\be
\hskip 1.0cm
\bigg[
( 1 - 2/r_{\Psi}^2 ) 
\frac{f^2}{\x\y} 
+
2\,(r_{\Psi}^2 f - 1)
\bigg]
\ .
\label{eq:dsigdt3} 
\ee
Here $\x$ and $\y$ are dimensionless, 
shifted Mandelstam variables
$\x \equiv t/m^2-1$ and $\y \equiv u/m^2-1$,
and $f$ is a dimensionless energy variable, which for general
masses is given by 
$f = (s - m_{\pi}^2 - M^2)/m^2 = -(\x + \y )$. 

There are several interesting features in these angular distributions which are 
relevant to PANDA. Note that the differential cross sections in 
Eqs.(\ref{eq:dsigdt1}-\ref{eq:dsigdt3}) are 
$t\leftrightarrow u$ ($\x\leftrightarrow \y$)
crossing symmetric, and that there are maxima in the 
forward and backwards directions, which give the largest values for 
the proton propagator functions 
$1/|t-m^2| = 1/m^2|\x|$ and $1/|u-m^2| = 1/m^2 |\y|$. A more
striking effect is the zero in the angular distribution of the final state 
$\pi^0 \eta_c$, which
follows from the odd $t\leftrightarrow u$ spatial symmetry of this 
amplitude; this implies a node at $t=u$, corresponding to 
$\theta = \pi/2$ in the $c.m.$~frame. In comparison the angular distributions of the 
other final states we consider also have minima at $t=u$, 
but are nonzero there. These rapidly varying angular 
distribution could be used to test for the presence of this signal over 
the presumably more slowly varying backgrounds. 

\subsubsection{Total cross sections}

In the massless pion limit 
the unpolarized total cross sections 
follow from 
integrating the results in Eqs.(\ref{eq:dsigdt1}-\ref{eq:dsigdt3}) 
over $t$. They may conveniently be written as 
functions of $s$ and $\beta = \sqrt{1-4m^2/s}$ (the $p$ or $\bar p$  
velocity in the $c.m.$ frame), and are given by

\be
\langle
\sigma
\rangle\Big|_{p\bar p \to \pi^0 \eta_c} 
=
2 \pi \alpha_{\pi}\alpha_{\Psi}\,
\frac{(s-M^2)}{s^2 \beta^2} \Big[ \tanh^{-1}\!\beta - \beta\, \Big]
\ee

\be
\hskip -1.35cm
\langle
\sigma
\rangle
\Big|_{p\bar p \to \pi^0 \chi_0}
=
2 \pi \alpha_{\pi}\alpha_{\Psi}\,
\frac{(s-M^2)}{s^2 \beta^2}\; \tanh^{-1}\!\beta
\ee

\bd
\hskip -2.65cm
\langle
\sigma
\rangle
\Big|_{p\bar p \to \pi^0 \chi_1}
=
2 \pi \alpha_{\pi}\alpha_{\Psi}\,
\frac{(s-M^2)}{s^2 \beta^2} \ \cdot
\ed
\be
\hskip 1.5cm
\bigg[\,
2(1-2/r{_\Psi}^2)\tanh^{-1}\!\beta\;
+ 
(s/M^2 -2)  \beta\;
\bigg].
\ee
The massless pion total cross section formulas for the 
final states 
$\pi^0 J/\psi$ and $\pi^0 \psi'$ 
are proportional to the $\pi^0 \chi_0$ result, as implied by
Eq.(\ref{eq:dsigdt2}); 
$
\sigma(J/\psi)
=
\sigma(\psi')
=
2 \sigma(\chi_0) 
$. 
We stress that these simple ratios only apply to the algebraic expressions; 
the actual numerical cross sections are not simply related, due to 
{\it a)} 
different kinematics, and 
{\it b)} the strongly state-dependent $g_{p\bar p \Psi}$ couplings, which are given in Table~\ref{tab:couplings}.

\subsection{Massive pions}

\subsubsection{Kinematics}

The results given in the previous section for massless pions are attractive
in their simplicity, and are useful for numerical estimates well above threshold.
However we find that the angular distributions near threshold are strongly dependent
on the pion mass (the massless case is a singular limit), and the difference in the location
of the threshold for massless versus massive pions leads to numerically important differences
in the predicted cross sections at energies relevant to PANDA. 
For these reasons we give detailed predictions for the differential and total cross sections
for these charmonium production processes in the case of general pion mass.

Since we will discuss some angular distributions in the {\it c.m.} frame, it is useful to have the 
relation between Mandelstam variables and proton $(p)$ and pion $(k)$ variables in this frame for general
masses. These relations are
\begin{eqnarray}
&E_p =& \frac{s^{1/2}}{2} 
\label{eq:E_p}
\\
&p =& \frac{s^{1/2}}{2}\; \Big(1-\frac{4m^2}{s}\Big)^{1/2}
\label{eq:p}
\\
&E_k =& \frac{s^{1/2}}{2}\; \Big(1-\frac{(M^2 - m_{\pi}^2)}{s}\Big)
\label{eq:E_k}
\\
&k =& \frac{s^{1/2}}{2} 
\bigg(1-\frac{2(M^2 + m_{\pi}^2)}{s} + \frac{(M^2 - m_{\pi}^2)^2}{s^2}\bigg)^{1/2}
\label{eq:k}
\end{eqnarray}

\subsubsection{Differential cross sections}

For general pion mass the differential cross sections 
$\{ \langle d\sigma/dt \rangle\} $ 
may be written as multiplier functions 
$\{F_{\Psi}\}$ 
times the massless pion formulas of Eqs.(\ref{eq:dsigdt1}-\ref{eq:dsigdt3}). 
These functions, expressed as power series in $r_{\pi} = m_{\pi}/m$, are
\begin{eqnarray}
&F_{\eta_c} =& 
1 - 
\frac{r_{\Psi}^2}{\x\y}
\, r_{\pi}^2
\label{eq:F1}
\\
&F_{\chi_0} =&  
1 - 
\frac{(r_{\Psi}^2-4)}{\x\y}  
\, r_{\pi}^2
\label{eq:F2}
\\
&F_{J/\psi, \psi'} =& 
1 +
\Big(
\frac{2(r_{\Psi}^2 + f)}{f^2}
-
\frac{(r_{\Psi}^2 + 2)}{\x\y}  
\Big) 
\, r_{\pi}^2
+
\, \frac{2}{f^2}
\, r_{\pi}^4 
\ .
\nonumber
\\
&& 
\label{eq:F3}
\end{eqnarray}
In the $\chi_1$ case it is simpler to give the full differential
cross section with nonzero pion mass directly. This result is

\bd
\hskip -2.5cm
\langle
\frac{d\sigma}{dt}
\rangle
\Big|_{p\bar p \to \pi^0 \chi_1} 
= 
\pi
\frac{\alpha_{\pi}\alpha_{\Psi}}{s (s-4m^2)}\, \cdot
\ed
\bd
\hskip -2.5cm
\Bigg\{
\bigg[
2\, (f/r_{\Psi}^2  - 1)
+
\frac{( 1 - 2 /r_{\Psi}^{2})f^2}{\x\y} 
\bigg]
\ed
\be
\hskip 0.5cm
+\; 
\bigg[
\frac{2}{r_{\Psi}^2} 
+ 
\frac{2 (r_{\Psi}^2-f-4)}{\x\y}
- 
\frac{(r_{\Psi}^2-4) f^2}{(\x\y)^2} 
\bigg] 
r_{\pi}^2
- 
\frac{2}{\x\y} r_{\pi}^4
\bigg\}
\ .
\label{eq:dsigdt4} 
\ee
Examples of these angular distributions will be shown in the section 
on numerical results.  

\subsubsection{Total cross sections}

In the case of general masses, inspection of the differential cross sections implied 
by Eqs.(\ref{eq:dsigdt1}-\ref{eq:dsigdt3}) and  
Eqs.(\ref{eq:F1}-\ref{eq:dsigdt4})
shows that the total cross sections may all be evaluated 
analytically in terms of integrals over $\x$, which are generically of the form

\be
{\cal I}_{m} = \int_{\x_0}^{\x_1} \frac{d\x}{(\x\y)^m}
\label{eq:Im_def} 
\ee
where $\y = -\x - f$.
The limits of integration are implied by the definition $\x = t/m^2 -1$, and 
in terms of proton $(p)$ and pion $(k)$ {\it c.m.}~frame energies and three-momenta are
\be
\x_{_{1\atop 0}}
= 
( m_{\pi}^2 - 2E_p E_k \pm 2 pk )/m^2. 
\ee
These limits may also be written in terms of masses and the Mandelstam variable $s$, 
using the relations given in Eqs.(\ref{eq:E_p}-\ref{eq:k}).
The explicit indefinite integrals required for our cross section calculations 
are 
\begin{eqnarray}
&{\cal I}_{0}(x) =& \x 
\label{eq:I0}
\\
&{\cal I}_{1}(x) =& \frac{1}{f} \ln\Big( \frac{\x +f}{\x} \Big)
\label{eq:I1}
\\
&{\cal I}_{2}(x) =& 
\frac{2}{f^3} \ln\Big( \frac{\x +f}{\x} \Big)
- \frac{1}{f^2} \bigg( \frac{1}{\x +f} + \frac{1}{\x} \bigg).
\label{eq:I2}
\end{eqnarray} 
In terms of these integrals, evaluated between the limits $\x_0$ and $\x_1$ 
(so that ${\cal I}_{m} \equiv {\cal I}_{m}(\x_1) - {\cal I}_{m}(\x_0)$), 
the total cross sections are 
\bd
\hskip -3.2cm
\langle
\sigma
\rangle_{p\bar p \to \pi^0 \eta_c } 
=
\frac{\pi \alpha_{\pi}\alpha_{\Psi}}{2}\,
\frac{m^2}{s(s-4m^2)}\, \cdot
\ed
\be
\hskip -2.0cm
\bigg[
- (4\, {\cal I}_{0} - f^2\, {\cal I}_{1} ) +
r_{\pi}^2 r_{\Psi}^2 \Big(4\, {\cal I}_{1} - f^2\, {\cal I}_{2}\Big)
\bigg]
\label{eq:sigtot_etac}
\ee

\bd
\hskip -2.5cm
\langle
\sigma
\rangle_{p\bar p \to \pi^0 (J/\psi, \psi' )} 
=
{\pi \alpha_{\pi}\alpha_{\Psi}}\,
\frac{m^2}{s(s-4m^2)}\, \cdot
\ed
\be
\bigg[\,
f^2\, {\cal I}_{1} 
+ r_{\pi}^2\Big(
2(r_{\Psi}^2 + f) {\cal I}_{1}
- (r_{\Psi}^2 + 2)\,f^2\, {\cal I}_{2}\Big)
+  
r_{\pi}^4
\, 2 {\cal I}_{1}
\bigg]
\label{eq:sigtot_Jpsi}
\ee

\be
\hskip 0.1cm
\langle
\sigma
\rangle_{p\bar p \to \pi^0 \chi_0 } 
=
\frac{\pi \alpha_{\pi}\alpha_{\Psi}}{2}\,
\frac{m^2 f^2}{s(s-4m^2)}\,
\bigg[\,
{\cal I}_{1} - r_{\pi}^2 \Big( r_{\Psi}^2  - 4\Big) {\cal I}_{2}
\bigg]
\\
\label{eq:sigtot_chi0}
\ee

\bd
\hskip -3.2cm
\langle
\sigma
\rangle_{p\bar p \to \pi^0 \chi_1} 
=
{\pi \alpha_{\pi}\alpha_{\Psi}}\,
\frac{m^2}{s(s-4m^2)}\, \cdot 
\ed
\bd
\hskip -2.5cm
\bigg[
\Big(\,
2(f/ r_{\Psi}^2-1) \, {\cal I}_{0} + f^2\, (1-2/r_{\Psi}^2)\, {\cal I}_{1}
\Big)\,
\ed
\be
+ 
r_{\pi}^2
\Big(\,
2 /r_{\Psi}^2 \, {\cal I}_{0} 
+
2 ( r_{\Psi}^{2} -f -4)\, {\cal I}_{1}
-
f^2\,  ( r_{\Psi}^{2} -4) \, {\cal I}_{2}
\Big)\,
-
r_{\pi}^4
2\, {\cal I}_{1} 
\bigg].
\label{eq:sigtot_chi1}
\ee
These cross sections will be evaluated numerically in the next section.

\section{Numerical Results}

\subsection{Estimating the $p\bar p \Psi $ coupling constants}

Numerical evaluation of our predictions for cross sections and related quantities
requires values for the model parameters. The hadron masses 
are of course well known, as is the NN$\pi$ coupling constant 
$g_{\pi} \equiv g_{pp\pi}$, 
which we take to be 13.5. The coupling strengths of the various 
charmonium resonances to $p\bar p$ 
however are not well established; this is a crucial issue for the PANDA project, which assumes 
that these couplings are sufficiently large to allow the accumulation of large charmonium event 
samples in $p\bar p$ annihilation.

Since the partial widths of several charmonium states to $p\bar p$ are now known experimentally,
one may estimate these
state-dependent charmonium-$p\bar p$ coupling constants $\{ g_{\Psi}\}$ by equating the measured
partial widths to the $O(g_{\Psi}^2)$ theoretical partial widths we calculate in our model,
assuming the same pointlike $p \bar p \Psi $ vertices that we used to calculate the 
$p\bar p \to \pi^0 \Psi$ cross sections. In terms of 
the $\alpha_{\Psi} = g_{p\bar p\Psi}^2/4\pi$ and the final $p$ velocity
$\beta = \sqrt{1-4m^2/M^2}$, these partial widths are
\begin{eqnarray}
&&
\Gamma(\eta_c \to p\bar p) 
\hskip 3.5mm
=  \alpha_{\eta_c} \beta M / 2
\label{eq:etac_ppbarwidth}
\\
\nonumber
\\
&&
\Gamma(J/\psi \to p\bar p) 
\hskip 0.5mm
=  \alpha_{J/\psi} \beta
( 1 + 2/r_{\Psi}^2 ) M / 3
\label{eq:Jpsi_ppbarwidth}
\\
\nonumber
\\
&&
\Gamma(\chi_0 \to p\bar p) 
\hskip 3mm
=  \alpha_{\chi_0} \beta^{\, 3} M / 2
\label{eq:chi0_ppbarwidth}
\\
\nonumber
\\
&&
\Gamma(\chi_1 \to p\bar p) 
\hskip 3mm
=  \alpha_{\chi_1} \beta^{\, 3} M / 3. 
\label{eq:chi1_ppbarwidth}
\end{eqnarray}
We use these relations and the measured branching fractions and total widths
to estimate values for the various charmonium-$p\bar p$ coupling constants; these 
results are shown in Table~I. Note that the couplings of the pseudoscalar and scalar states 
$\eta_c$ and $\chi_0$ to $p\bar p$ are an order of magnitude larger that the couplings of the 
vector and axial vector states. 

\begin{table}[h]
\begin{center}
\begin{tabular}{|c|c|c|c|}
\hline
State $\Psi$ 
& B$_{\Psi\to p\bar p}$ 
& $\Gamma^{tot.}_{\Psi}$[MeV]
& $10^3 \cdot g_{p\bar p \Psi}$ 
\\ 
\hline
\
$\eta_c$ 
& $(1.3\phantom{0} \pm 0.4\phantom{0}) \cdot 10^{-3}$               
& $25.5\phantom{000} \pm 3.4\phantom{000}$
& $19.0\phantom{0} \pm 3.2\phantom{00} $
\\
$J/\psi$ 
& $(2.17 \pm 0.08) \cdot 10^{-3}$               
& $\phantom{0}0.0934 \pm 0.0021$
& $\phantom{0}1.62 \pm 0.03 \phantom{0} $
\\
$\psi'$ 
& $(2.65 \pm 0.22) \cdot 10^{-4}$               
& $\phantom{0}0.337\phantom{0} \pm 0.013\phantom{0}$
& $\phantom{0}0.97 \pm 0.04 \phantom{0} $
\\
$\chi_0$ 
& $(2.24 \pm 0.27) \cdot 10^{-4}$               
& $10.4\phantom{000} \pm 0.7\phantom{000}$
& $\phantom{0}5.42 \pm 0.37 \phantom{0} $
\\
$\chi_1$ 
& $(6.7\phantom{0} \pm 0.5\phantom{0}) \cdot 10^{-5}$               
& $\phantom{0}0.89\phantom{00} \pm 0.05\phantom{00}$
& $\phantom{0}1.03 \pm 0.07 \phantom{0} $
\\
$\chi_2$ 
& $(6.6\phantom{0} \pm 0.5\phantom{0}) \cdot 10^{-5}$               
& $\phantom{0}2.06\phantom{00} \pm 0.12\phantom{00}$
& $ - \phantom{0} $
\\
\hline
\end{tabular}
\caption{Charmonium-$p\bar p$ coupling constants, estimated 
from the measured branching fractions and total widths \cite{PDG} and the 
$\Gamma_{\Psi \to p\bar p}$ width formulas of 
Eqs.(\ref{eq:etac_ppbarwidth}-\ref{eq:chi1_ppbarwidth}) (see text).}
\label{tab:couplings}
\end{center}
\end{table}

\subsection{Total Cross Sections}
\begin{figure}[h]
\includegraphics[width=0.9\linewidth]{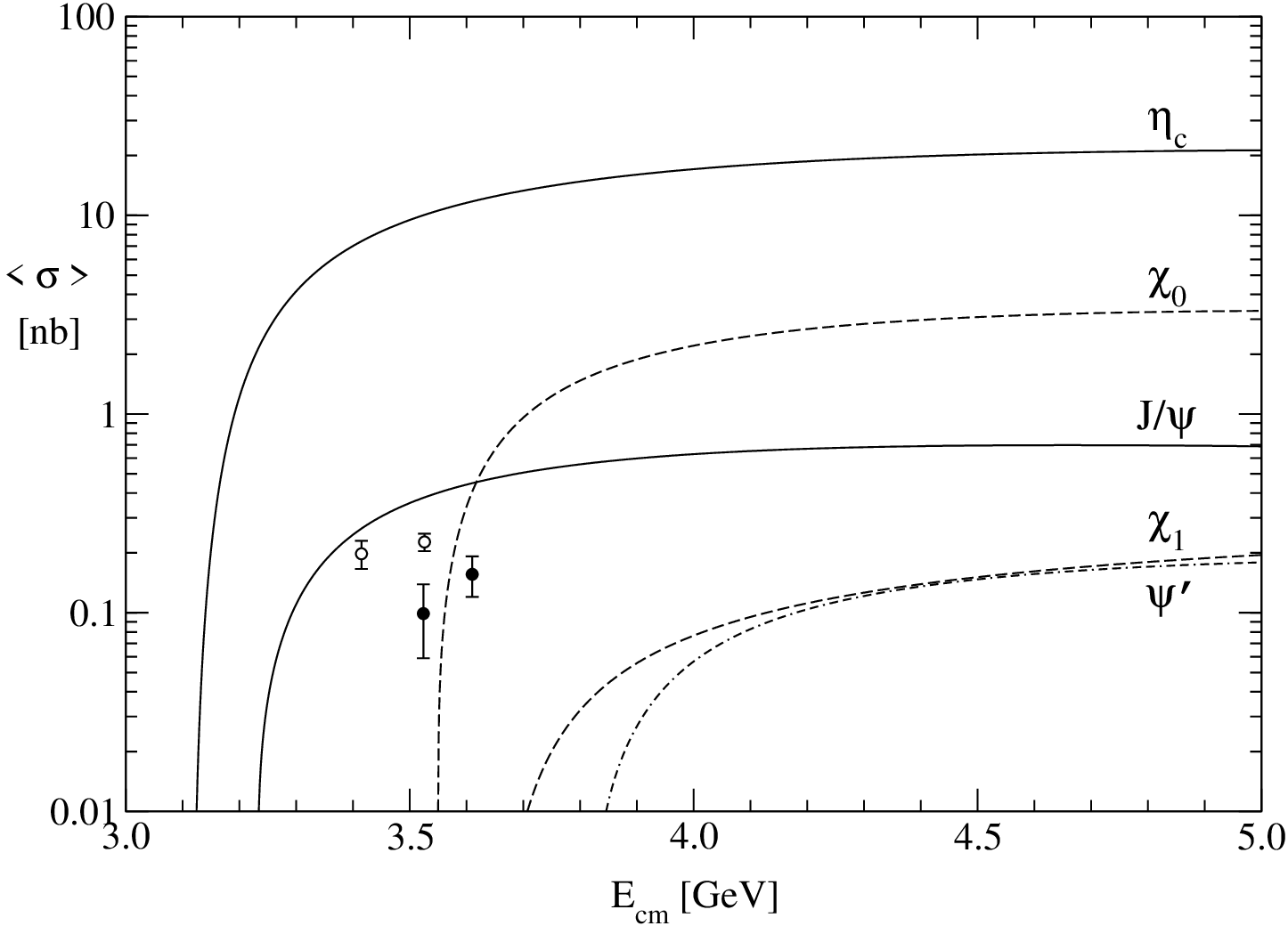}
\caption{Predicted unpolarized total cross sections for the processes
$p \bar p \to \pi^0 \Psi$, where $\Psi = \eta_c, J/\psi, \chi_0, 
\chi_1$ and $\psi'$. The data points are Fermilab measurements of the
cross section for $p \bar p \to \pi^0 J/\psi$, from E760 \cite{Armstrong:1992ae} (filled) 
and E835 \cite{Joffe:2004ce} (open).
There are additional experimental results for this reaction from E835 \cite{Andreotti:2005vu} 
which have not yet been published as a physical cross section.} 
\label{fig:csecs}
\end{figure}

We will now evaluate the total unpolarized cross sections   
$\langle \sigma \rangle (p \bar p \to \pi^0 \Psi)$ 
for the five cases $\Psi = \eta_c, J/\psi, \chi_0,
\chi_1$ and $\psi'$, using the formulas given in 
Eqs.(\ref{eq:sigtot_etac}-\ref{eq:sigtot_chi1}). The masses 
assumed are 2006 PDG values \cite{PDG} rounded to 0.1~MeV;
$m_{\pi^0}  = 0.1350$~GeV,
$m_{p}  = 0.9383$~GeV
$M_{\eta_c} = 2.9804$~GeV,
$M_{J/\psi} = 3.0969$~GeV,
$M_{\chi_0} = 3.4148$~GeV,
$M_{\chi_1} = 3.5107$~GeV
and
$M_{\psi' } = 3.6861$~GeV. 
The $pp\pi^0$ coupling constant is taken to be
$g_{\pi} = 13.5$, and the $p\bar p\Psi$ couplings are given in 
Table~\ref{tab:couplings}. 
The resulting cross sections are shown in
Fig.\ref{fig:csecs}.
\vskip 0.5cm

Evidently these cross sections share a rapid rise above threshold, followed by
a slowly varying ``plateau" region above E$_{cm} \approx 4$~GeV. 
The most remarkable feature may be the wide variation in the 
plateau values with the charmonium state $\Psi$. 
The largest cross section by far is for
$\eta_c$ production, which is roughly a factor of 30 larger than the $J/\psi$ cross section. 
Second is the $\chi_0$ cross section, roughly 5 times larger than $J/\psi$. Since the 
$\chi_0$ has a radiative branching fraction to $\gamma J/\psi$ of $35.6\pm 1.9 \%$, this implies that
$J/\psi$ production through the radiative process $p \bar p \to \pi^0 \chi_0$, $\chi_0 \to \gamma J/\psi$ 
is comparable to the direct process $p \bar p \to \pi^0 J/\psi$. Finally, the
$\chi_1$ and $\psi'$ cross sections are smaller than the $J/\psi$ by about a factor of 5-10.
(For the $\psi'$ to $J/\psi$ ratio this is reminiscent of the $12\%$ rule 
for a radial excitation.)
These results suggest that the associated production of charmonium and charmonium hybrids 
with certain quantum numbers is strongly enhanced in $p\bar p$ production, and that 
$J^{PC} = 0^{-+}$ and $0^{++}$ states may be the most favored. 

\subsection{Angular distributions}

The angular distributions predicted by this model of $p \bar p \to \pi^0 \Psi$ are especially interesting,
since they are far from isotropic and may be important for design of the PANDA detector. Two representative 
cases for $\langle d\sigma / d\Omega\rangle$ ($c.m.$~frame) are shown in the figures; both show angular distributions normalized
to unity in the forward direction, with E$_{cm}$ in steps of 0.2~GeV up to a cutoff of  E$_{cm} = 5.0$~GeV.
The contours at E$_{cm} = 4.0$~GeV and E$_{cm} = 5.0$~GeV are highlighted.  

\vskip 0.8cm
\begin{figure}[h]
\includegraphics[width=0.8\linewidth, angle=0]{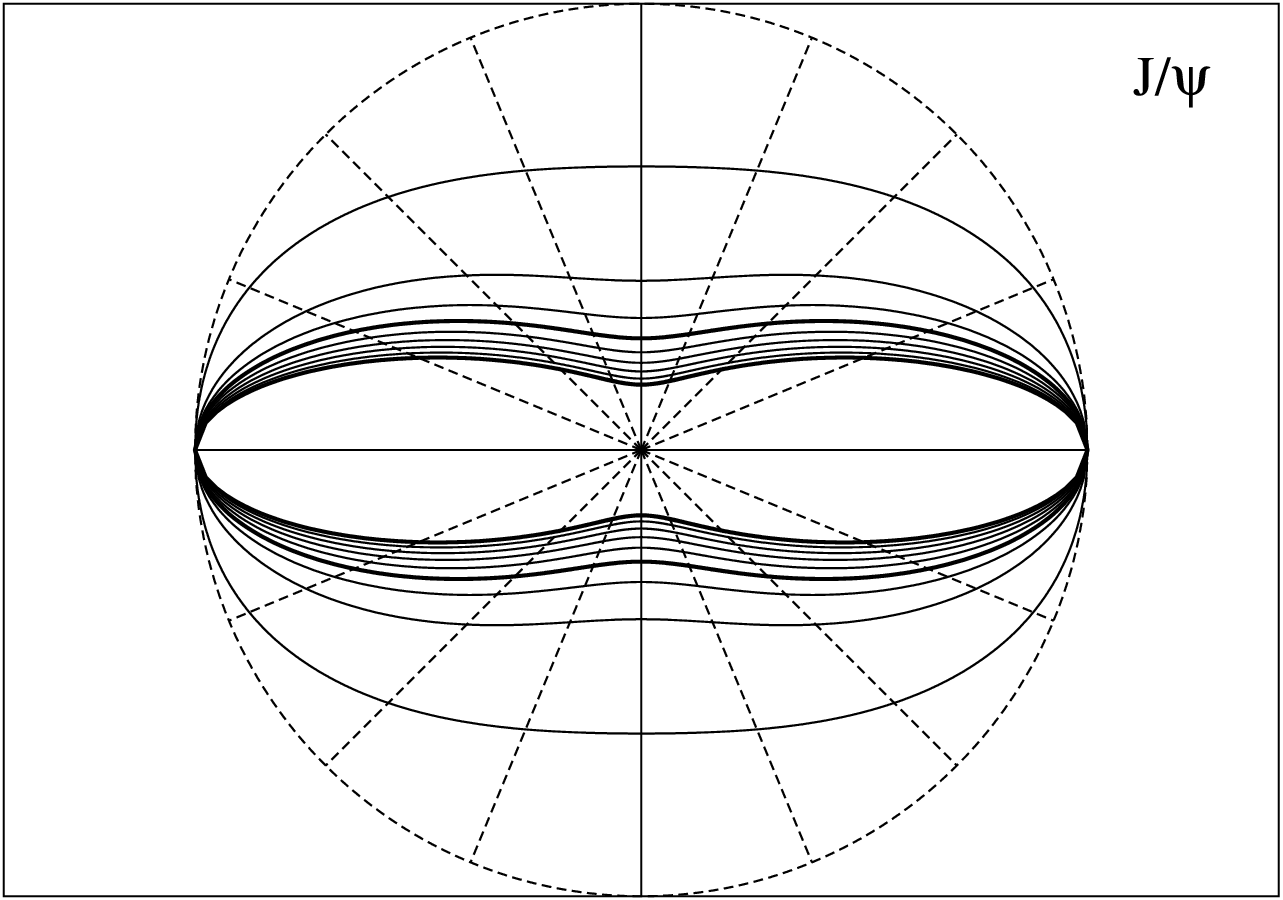}
\vskip -0.2cm
\caption{Predicted $c.m.$~frame unpolarized angular distribution 
$\langle d\sigma / d\Omega\rangle $ for the process $p \bar p \to \pi^0 J/\psi$, 
normalized to the forward intensity, for E$_{cm} = 3.4-5.0$ GeV
in steps of 0.2 [GeV].}
\label{fig:W_Jpsi}
\end{figure}

\vskip 0.1cm
\begin{figure}[h]
\includegraphics[width=0.8\linewidth, angle=0]{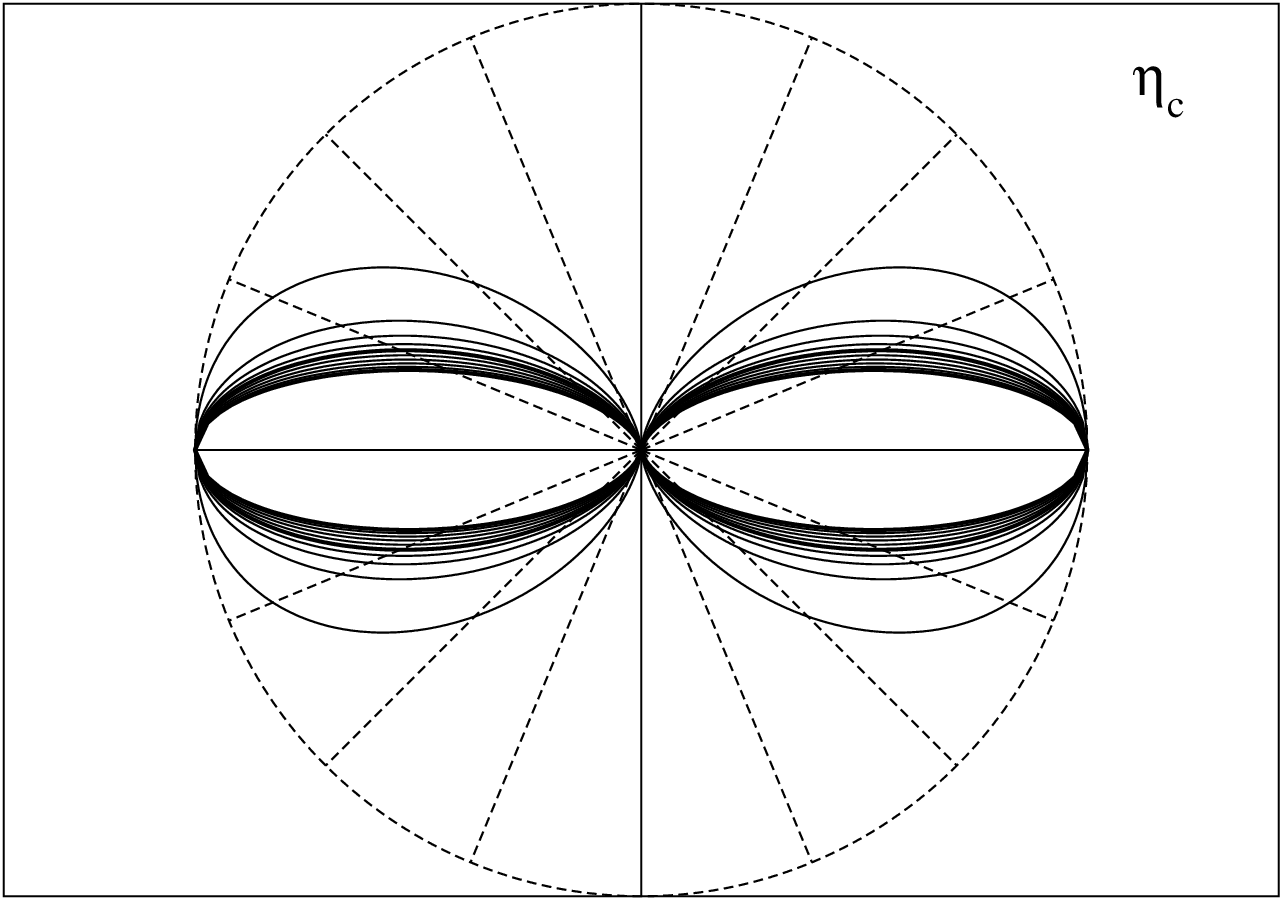}
\vskip -0.2cm
\caption{The corresponding angular distribution for the reaction $p \bar p \to \pi^0 \eta_c$,
for E$_{cm} = 3.2-5.0$ GeV in steps of 0.2~[GeV].}
\label{fig:W_etac}
\end{figure}

The first angular distribution, in Fig.\ref{fig:W_Jpsi}, shows the prediction of 
Eqs.(\ref{eq:dsigdt2},\ref{eq:F3}) for the reaction
$p \bar p \to \pi^0 J/\psi$. This distribution is isotropic as we approach threshold, but with increasing 
E$_{cm}$ evidently becomes strongly forward- and backward-peaked. (All the unpolarized cross sections we 
consider are front-back symmetric.)

The second angular distribution is for the reaction $p \bar p \to \pi^0 \eta_c$, which is predicted to be 
the largest associated charmonium production cross section by a considerable margin. This angular distribution,
taken from Eqs.(\ref{eq:dsigdt1},\ref{eq:F1}), is shown in Fig.\ref{fig:W_etac}. 
(The same conventions are used as in the previous figure.)
This differential cross section shows a similar forward- and backward-peaking with increasing E$_{cm}$, and 
has in addition a very characteristic node at $t=u$, which is $\theta = 90^o$ in the $c.m.$~frame. The large
overall scale of this cross section suggests that the detection of unusual modes such as $p\bar p \to 4\gamma$ 
may be feasible (with both the $\pi^0$ and $\eta_c$ decaying to $\gamma\gamma$).

\section{Summary and Future}

In this paper we have developed a hadronic ``pole" model of charmonium production with 
an associated pion in proton-antiproton collisions. Given the Feynman diagrams of 
Fig.\ref{fig:diags} we derived closed-form analytic results for the unpolarized 
differential and total cross sections for the processes $p\bar p \to \pi^0 \Psi$, 
where the quantum numbers 
considered for the charmonium state $\Psi$ were $0^{-+}, 0^{++}, 1^{--}$ and $1^{++}$. 
We quoted results for both massless and massive pions.
 
Besides the hadron masses this 
model has as free parameters the known $g_{pp\pi^0}$ and the coupling 
$g_{p\bar p \Psi}$ of the specified
charmonium state $\Psi$ to $p\bar p$. Here we used the experimental partial 
widths of light charmonium states to $p\bar p$ to estimate the coupling constants
$\{ g_{p\bar p \Psi}\} $. Given this information we numerically evaluated   
the differential and total cross sections for these reactions. In particular we gave 
numerical predictions for the total cross section for the cases 
$\Psi = \eta_c, J/\psi, \chi_0, \chi_1$ and $\psi'$
(Fig.\ref{fig:csecs}),
and for the differential cross sections
in two representative cases, $\Psi = J/\psi$ 
(Fig.\ref{fig:W_Jpsi})
and $\eta_c$ (Fig.\ref{fig:W_etac}). 
The two published data points
from E760 for the $p\bar p \to \pi^0 J/\psi$ total cross section near 3.5~GeV 
(Fig.\ref{fig:csecs})
suggest that our results overestimate the cross section at this E$_{cm}$ by about a 
factor~of~2.

The total cross sections predicted for the other reactions are especially interesting. The 
reaction $p\bar p \to \pi^0 \eta_c$ is predicted to have the largest cross section by a 
considerable margin, followed by the $\chi_0$. The two smallest cross sections are 
predicted for the $\chi_1$ and $\psi'$. These results suggest that 
$J^{PC} = 0^{-+}$ and $0^{++}$ charmonium states may be the most copiously produced 
in the PANDA experiment. A comparison of the $J/\psi$ and $\psi'$ suggests that the 
first radial excitation cross sections are suppressed relative to the ground states 
by about a factor of 5-10 in the relevant PANDA energy regime. 

Several future developments related to this type of model appear especially
interesting. The important question of the size of the associated production cross sections
of other charmonium states can be answered in this model given their
$p\bar p$ branching fractions, so $B_{p\bar p}$ is a very important measurements for the 
charmonium states we have not considered here. Similarly, the cross sections for emission 
of a charmonium state $\Psi$ accompanied by other light mesons $\{m\}$ can be studied; 
this may also be used to clarify poorly known $ppm$ couplings. 
The importance of initial and final 
polarizations, which may be accessible at PANDA, can easily be addressed in this model. 
An important effect which has not been 
included is the contribution of intermediate N$^*$ resonances to the 
$p\bar p \to \pi^0 \Psi$ transition amplitudes; the ${\rm NN}^*\Psi$
couplings can be extracted from the corresponding $\Psi \to {\rm NN}^*$ decays. 
Finally, as our approximation of pointlike
hadron vertices is unrealistic well above threshold, it would be very interesting to 
study the effect of plausible $pp\pi^0$ and $p\bar p\Psi$ form factors on our results.  

\section{Acknowledgements}

We are happy to acknowledge useful communications with
D.Bettoni, C.Patrignani, T.Pedlar and K.Seth regarding experimental results for
associated charmonium production cross sections at Fermilab. We also gratefully 
acknowledge support from K.Peters at GSI in Summer 2006, where this work was initiated.
This research was supported in part by 
the U.S. National Science Foundation through grant NSF-PHY-0244786 at the
University of Tennessee, and the U.S. Department of Energy under contract
DE-AC05-00OR22725 at Oak Ridge National Laboratory.

\vfill\eject

\end{document}